\documentclass{Rinton-P9x6}

\usepackage{amsmath,amsfonts,amssymb}

\newcommand{\by}{\mathbf {y}}
\newcommand{\bp}{\mathbf {p}}

\begin{document}

\title{Geometry Dependence of Casimir Forces beyond the Proximity
  Approximation}

\author{Thorsten Emig}

\address{Institut f\"ur Theoretische Physik, Universit\"at zu
K\"oln, Z\"ulpicher Stra\ss e 77, D-50937 K\"oln, Germany}

\maketitle

\abstracts{Casimir interactions between macroscopic objects are
  strongly influenced by their geometrical features as shape and
  orientation as well as by their material properties. The effect of
  geometry is commonly obtained from the proximity approximation (PA).
  Here we present a path integral quantization for the electromagnetic
  field in the presence of deformed metallic surfaces. From the
  resulting effective action the Casimir force between the surfaces
  can be calculated without the PA. For corrugated surfaces the force
  is obtained both perturbatively for small deformations and by a
  numerical approach for general deformation amplitudes.  For general
  dielectric materials with flat surfaces a path integral based
  derivation of the Lifshitz theory is outlined, pointing towards a
  possible approach to study the combined effect of deformations and
  material properties.}

\section{Introduction and Summary}
\label{sec:intro}

Casimir's seminal prediction of an universal attractive force between
two uncharged metallic plates has been confirmed recently in the
distance range from 100nm to several micrometer with increasing
accuracy in a number of beautiful high precision
experiments.\cite{Lamoreaux97,Mohideen+98,Lambrecht+00,Ederth00,Chan+01,Bressi+02,Decca+03}
High accuracy could be only achieved by a careful inclusion of the
corrections to Casimir's ideal result from finite conductivity of the
materials, surface roughness and non-zero temperature.\cite{Bordag+01}
Except one, \cite{Bressi+02} the above cited experiments deviate from
the flat-plate geometry in that they use a plane-sphere configuration
in order to avoid the difficulty of maintaining a constant plate
separation over the entire surface area.  The plane-sphere
configuration is the standard situation where the proximity (or
Derjaguin) approximation (PA) \cite{Derjaguin34} is commonly employed
to calculate geometry induced changes of the force.  Consider a
non-planar surface above a planar surface where their local distance,
measured normal to the planar surface, is given by $H+h(\by_\|)$ with
$\by_\|$ the 2D in-plane coordinates, i.e., $h(\by_\|)=0$ corresponds
to two parallel flat plates.  The PA assumes that the Casimir energy
can be computed as the sum of local contributions between {\it flat}
surface elements at their local distance, yielding the total energy
per unit surface area
\begin{equation}
\label{eq:pa}
{\cal E}_{\rm PA}=\int d^2 \by_\| \, {\cal E}_0[H+h(\by_\|)],
\end{equation}
where ${\cal E}_0(H)=-(\pi^2/720) (\hbar c/H^3)$ is the Casimir energy
per unit area between two plane surfaces. This approximation is
ambiguous since one could just as well measure the local distances
normal to the curved surface, yielding a different result. Below, we
will discuss this point further when comparing the PA to our
perturbative and numerical results.  The PA is expected to hold if
both the local surface curvature radii are much larger than the local
distance and strongly non-parallel surface elements are at larger
separations than more parallel ones.\cite{Gies+03} A conceptual
different approximation is the pairwise summation of renormalized
retarded van der Waals forces.\cite{Bordag+01} For the geometry
considered here it yields exactly the PA result of
Eq.~(\ref{eq:pa}).\cite{Emig+02} For the plane-sphere geometry used in
experiments, one can set $h(\by_\|)=\by_\|^2/2R$ with radius of
curvature $R$, yielding the approximate force $F_{\rm PA}=2\pi R {\cal
  E}_0(H)$.  The latter formula is used in experiments, and there it
is justified for that $R$ is much larger than $H$.\footnote{A
  numerical scalar field analysis yields sizable deviations from the
  PA for $R \lesssim 50 H$.\cite{Gies+03}}

For more general geometries the collective nature of the Casimir force
suggests it to have a non-trivial and unexpected dependence on the
shape of the interacting objects. Whereas the van der Waals force
between electrically polarizable particles is always attractive, even
the sign of the Casimir force is geometry dependent. There is little
intuition for the value of the sign as demonstrated by the repulsive
force for a thin conducting shell.\cite{Boyer74,Balian+78} Due to the
importance of Casimir forces in many fundamental and applied contexts,
it is highly desirable to obtain a better understanding of its strong
geometry dependence, including distinctions from pairwise additive
interactions and possibly repulsive forces between {\it disconnected}
surfaces.  A promising and experimentally testable route to this end
is via modifications of the parallel plate geometry.\cite{Emig+01} In
an experimental search for novel shape dependencies, Mohideen {\it et
  al.} measured the force between a sphere and a sinusoidally
corrugated plate with a wave length being larger than the studied
range of sphere-plate separations.\cite{Roy+99} Their results showed
clear deviations from the PA prediction. While it has been suggested
that lateral forces caused this deviations\cite{Klimchitskaya+00},
our results indicate a much stronger sensitivity to geometry at
smaller corrugation wave lengths. 

\begin{figure}[h]
\begin{center}
  \epsfxsize=28pc \epsfbox{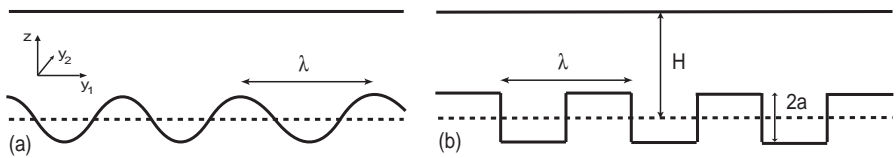}
\caption{Geometry composed of a flat and a corrugated surface which is studied 
  by perturbation theory (a) and a numerical approach (b). For this two
  cases the corrections to the proximity approximation scale
  differently to zero for increasing $\lambda$.
  \label{fig:1}}
\end{center}
\end{figure}

We demonstrate that a path integral quantization of the
electromagnetic field subject to appropriate boundary conditions is a
powerful method both for perturbative \cite{Emig+01,Emig+02} and
numerical \cite{Emig03} computations of Casimir interactions. While
the well-known boundary conditions for ideal metals are local, we
derive non-local boundary conditions for general dielectric materials
which reproduce the standard Lifshitz theory in the limit of flat
surfaces. For the force between deformed surfaces of ideal metals we
find the following results. For general uniaxial deformations we
obtained the Casimir force perturbatively to second order in the
deformation amplitude $a$.  At this order, significant deviations from
the PA result are found if the deformation wavelength $\lambda$ is of
the same order or smaller than the mean surface distance $H$. We apply
the perturbative results to sinusoidally corrugated surfaces, see
Fig.~\ref{fig:1}(a). The deformation induced change in the force shows
for $\lambda \ll H$ a slower decay $\sim H^{-5}$ compared to the
$H^{-6}$ behavior from the PA. However, the perturbative approach is
limited to the cases where $a$ is smaller than both $H$ and $\lambda$.
This can be understood by comparing the perturbative approach to the
multiple scattering approach of Balian and Duplantier\cite{Balian+78}
which yields the Casimir energy as a sum over waves propagating along
closed paths which scatter at an even number of positions on the
surfaces.  Each path contributes with a geometric factor which
measures at each position the scattering direction relative to the
local surface normal. For a series of scatterings on the same surface
with all scattering directions being almost perpendicular to the
surface normal the geometric factor is small. Our perturbative
approach allows in principle for paths with an arbitrarily large
number of scatterings. However, it corresponds to an expansion of the
geometric factor to second order in the surface deformations. This
amounts to the restriction to two successive scatterings on one
surface before the wave propagates to the other surface and
experiences for the rest of its path only scatterings normal to the
surface. This expansion is only justified for surface geometries where
$a$ is smaller than $\lambda$ so that the scattering directions lie
nearly in the mean surface plane.  The restriction can be bypassed by
an exact evaluation of the path integral by an algorithm which was
developed for corrugated surfaces.\cite{Emig03} This numerical
approach is applied to the rectangular grating of Fig.~\ref{fig:1}(b).
It confirms the crossover of the force change from a decay $\sim
H^{-6}$ to $\sim H^{-5}$ at $H\simeq \lambda$. In the limit of
$\lambda \lesssim a$ strong corrections to the perturbative results
are found.  For $\lambda \to 0$ the force corresponds to that
of two planar plates at a reduced distance $H-a$ and provides an upper
bound to the force at fixed $H$. For large $\lambda$ both the
perturbative and the numerical results approach the PA prediction.
However, we find that corrections to the PA decay slower with
increasing $\lambda/a$ for the profile with edges [Fig.\ref{fig:1}(b)]
as compared to the smooth profile [Fig.\ref{fig:1}(a)]. For this
finding we provide an interpretation in terms of classical ray optics
\cite{Jaffe+03}.

\section{Path Integral Approach}

Due to the lack of space we give only a very brief account of the path
integral approach\cite{Li+91} and focus instead on explicit results
for corrugated surfaces. For the uniaxial geometries considered here,
the electromagnetic field can be split into transversal electric (TE)
and magnetics (TM) modes which can both be described by a scalar field
$\Phi$. After a Wick rotation to imaginary time, the field is
quantized using the Euclidean action
\begin{equation} \label{eq:action}
S[\Phi] \, = \, \frac{1}{2} \int d^4 X \, (\nabla
\Phi)^2 .
\end{equation}
The change in the ground state energy due to the presence of
boundaries depends on the type of modes. TM modes are described by 
Dirichlet boundary conditions, $\Phi|_S=0$, and TE modes correspond to
Neumann boundary conditions, $\partial_{\bf n}\Phi|_S=0$, where
$\partial_{\bf n}$ is the normal derivative on the surface pointing
into the space between the plates. The Casimir interaction between the
two surfaces due to TM modes can be obtained from the restricted
partition function
\begin{equation}
\label{eq:Z}
{\cal Z_{\rm TM}}={\cal Z}_0^{-1} \int {\cal D} \Phi \prod_{\alpha=1}^2 
\prod_{X_\alpha} 
\delta[\Phi(X_\alpha)] e^{-S[\Phi]/\hbar},
\end{equation}
where ${\cal Z}_0$ is the partition function in the absence of
boundaries. For TE modes one obtains ${\cal Z}_{\rm TE}$ by replacing
the argument of the delta-function by $\partial_{\bf
  n}\Phi(X_\alpha)$. The position of the surfaces are parameterized in
the 4D Euclidean space by $X_1=[\by,h(\by_\|)]$, $X_2=[\by,H]$,
$\by=(ict\equiv y_0,\by_\|)$, where $H$ is their mean distance, and
the upper surface is planar. The partition functions can be computed
by expanding the delta-functions as $\prod_{X_\alpha}
\delta[\Phi(X_\alpha)] = \int {\cal D}\psi_\alpha
\exp[i\int_{S_\alpha} dX_\alpha \psi_\alpha \Phi]$ with an auxiliary
field $\psi_\alpha$ defined on each surface. Via the relation $F=
\hbar c/(AL) \partial_H \ln {\cal Z}$, with $L$ the Euclidean system
size in time direction, one finds for the force $F$ per unit area the
result
\begin{equation}
\label{eq:force}
F=-\frac{\hbar c}{2AL} {\rm Tr}\left( M^{-1} \partial_H M \right).
\end{equation}
This formula holds separately for both types of modes. For TM and TE
modes, respectively, the $2\times 2$ matrix kernel read
\begin{subequations}
\label{eq:kernels}
\begin{eqnarray}
M^{\text{TM}}_{\alpha \beta}({\bf y},{\bf y}') \, & = & \,
G[X_{\alpha}({\bf y})-X_{\beta}({\bf y}')] \, \\
M^{\text{TE}}_{\alpha \beta}({\bf y},{\bf y}') \, & = & \,
\partial_{n_\alpha(\by_\|)} \partial_{n_\beta(\by'_\|)}
G[X_{\alpha}({\bf y})-X_{\beta}({\bf y}')] \, 
\end{eqnarray}
\end{subequations}
with $G({\by},z)=(\by^2+z^2)^{-1}/4\pi^2$ the free Green's function.
Below, we will evaluate the force from Eq.~(\ref{eq:force}) both
perturbatively in $h(\by_\|)$ and numerically.

So far we applied the path integral approach only to ideal metal
surfaces.  If one assumes a splitting into TE and TM modes, arbitrary
dielectric surfaces can be described by a path integral with {\it
  non-local} boundary conditions for the scalar field
\cite{Buescher+03}. Planar surfaces with dielectric function
$\epsilon(\omega)$ are described by a mixed boundary condition which
reads in momentum $(k_0,{\bf k}_\|)$ representation
\begin{equation}\label{eq:scalar_LBC}
\left[1-\Gamma \,\partial_{\bf n}\!\right]\Phi|_{S}
\:=\:0
\end{equation}
with $\Gamma=[\epsilon(ik_0)k_0^2+{\bf k}_\|^2]^{-1/2}$ and
$\Gamma=\epsilon(ik_0)[\epsilon(ik_0)k_0^2+{\bf k}_\|^2]^{-1/2}$ for
TM and TE modes, respectively.  In the limit of ideal metals,
$\epsilon \to \infty$, this condition reduces to the above treated
Dirichlet and Neumann boundary condition for TM and TE modes,
respectively. The condition of Eq.~(\ref{eq:scalar_LBC}) together with
the action of Eq.~(\ref{eq:action}) provide a simple
derivation\cite{Buescher+03} of the standard Lifshitz
theory.\cite{Lifshitz56} More interestingly, the application of more
general non-local boundary conditions to gauge field path integrals
might prove useful for studying correlations between geometrical
and material dependencies of Casimir forces.

\section{Perturbative Results}

The logarithm of the partition function of Eq.~(\ref{eq:Z}) can be
expanded in the height profile.\cite{Li+91} With the choice that $\int
d^2 \by_\| h(\by_\|) =0$ there is no contribution at linear order in
$h(\by_\|)$. At second order the $H$-dependent part of $\ln {\cal Z}$
has the form
\begin{equation}
  \label{eq:gen-pert}
  \delta^{(2)}{\cal Z}=
\frac{\pi^2 L}{240 H^5} \!\int d^2\by_\| \, h^2(\by_\|) - 
\frac{L}{4} \!\int \! d^2\by_\| \!\int \! d^2\by'_\| \, K(|\by_\|-\by'_\||)
\left[ h(\by_\|) - h(\by'_\|) \right]^2
\end{equation}
with a kernel $K$ which is different for TM and TE
modes.\cite{Emig+02} From this general result the force for the
sinusoidal geometry of Fig.~\ref{fig:1}(a) can be computed. Assuming
$h(\by_\|)=a\cos(2\pi y_1/\lambda)$, the
total force from TM and TE modes assumes the form
\begin{equation}
\label{eq:pt-general}
F = F_{\rm flat} \left[ 1+ G\left(\frac{H}{\lambda}\right) 
\left(\frac{a}{H}\right)^2 + {\cal O}\left(a^3\right)\right]
\end{equation}
with $F_{\rm flat}=-(\pi^2/240) \hbar c/H^4$ the force per unit area
between two flat plates. Although the full form of the function
$G(H/\lambda)$ is available\cite{Emig+01,Emig+02}, we focus here on
two limiting cases. For small and large corrugation lengths $\lambda$,
one finds
\begin{equation}
  \label{eq:F-small-lam}
  F/F_{\rm flat} = 1 + \frac{8\pi}{3} \frac{a^2}{\lambda H} 
\,\,\text{for $\lambda \ll H$},\quad F/F_{\rm flat} = 1 + 5 \frac{a^2}{H^2}
\,\,\text{for $\lambda \gg H$}.
\end{equation}
Therefore, for small $\lambda$ the change of the force decays only
like $H^{-1}$ as compared to $\sim H^{-2}$ in the previously known
limit of large $\lambda$ for which the PA is expected to hold. The
divergence for small $\lambda$ is an artifact of the perturbative
approach as we will show below. Next we compare these results to the
predictions of the PA. Depending on which surface this approximation
is based, one obtains from Eq.~(\ref{eq:pa}) to second order in $a$
for the force
\begin{equation}
\label{eq:PA-cos}
F_{\rm PA,flat}/F_{\rm flat}= 1+ 5\frac{a^2}{H^2}, \quad
F_{\rm PA,corr}/F_{\rm flat}= 1+ 5\frac{a^2}{H^2} - 
3 \pi^2 \frac{a^2}{\lambda^2}
\end{equation}
for the flat and corrugated plate based PA, respectively. As expected,
the later result is smaller due to the increased surface separations
when measured normal to the corrugated plate. For $\lambda \to \infty$
both results coincide and agree also with the perturbative finding. It
is instructive to study how the difference between the full
perturbative result of Eq.~(\ref{eq:pt-general}) and the PA decays to
zero at large $\lambda$. For this difference we find the following
expressions which have opposite sign,
\begin{equation}
  \label{eq:diff-to-PA}
\frac{F-F_{\rm PA,flat}}{F_{\rm flat}}= \left( \frac{4\pi^2}{3}-20
\right) \frac{a^2}{\lambda^2}, \quad
\frac{F-F_{\rm PA,corr}}{F_{\rm flat}}= \left( \frac{13\pi^2}{3}-20 
\right) \frac{a^2}{\lambda^2}.
\end{equation}
Thus, the flat surface based PA overestimates the force while the
corrugated surface based PA yields a force which is too small. This
underlines the ambiguity of the PA, even for large $\lambda$ or small
surface curvature. Similar observations have been made for a
sphere-plane geometry.\cite{Gies+03} However, even the exponent of
$a/\lambda$ can depend on details of the corrugated surface. As we
will show below, for the rectangular corrugation of
Fig.~\ref{fig:1}(b) the decay towards the PA result is slower as for
the present geometry. In the next section we provide an explanation of
this strong geometry sensitivity in terms of geometric optics.

\section{Exact Numerical Results}

We have seen that the perturbative approach is limited to smooth
surface profiles where the deformation amplitude $a$ sets the smallest
of the geometrical length scales. The condition for the applicability
of perturbation theory is that two arbitrary points on the deformed
surface are connected by a vector which is nearly parallel to the
reference plane. This is certainly not true for the profile of
Fig.~\ref{fig:1}(a) if $\lambda \ll a$ but is more generally violated
if the profile has vertical segments as it is the case for the
rectangular corrugation of Fig.~\ref{fig:1}(b). For the latter
profile, the perturbative expression of Eq.~(\ref{eq:gen-pert}) indeed
diverges. However, the general result of the path integral approach,
Eq.~(\ref{eq:force}), can be used for a precise numerical computation
of the force even in cases where perturbation theory fails. For
periodic profiles one can apply the following approach.\cite{Emig03}
First, the Fourier transformed kernel $M(\by,\by')$ of
Eq.~(\ref{eq:kernels}) is decomposed into contributions from momenta
which are integer multiples of $2\pi/\lambda$, i.e.,
\begin{equation}
\label{eq:decomp}
M({\bf p},{\bf q})=\sum_{m=-\infty}^\infty  N_m(p_\perp,p_1) 
\, (2\pi)^3 \, \delta({\bf p}_\perp+{\bf q}_\perp) 
\, \delta(p_1+q_1+2\pi m/\lambda)
\end{equation}
with $\bp_\perp = (p_0,p_2)$. The $2\times 2$ matrices $N_m$ are
completely determined by a given uniaxial surface profile $h(y_1)$.
Using Eq.~(\ref{eq:force}), the force per unit area can be then
obtained from
\begin{equation}
\label{eq:F-non-pert}
F=-\frac{\hbar c}{4\pi^2}\int_0^\infty dp_\perp p_\perp 
\int_0^{\pi/\lambda} dp_1 \, g(p_\perp,p_1),
\end{equation}
with
\begin{equation}
\label{eq:fct-g}
g(p_\perp,p_1)={\rm tr} \left( B^{-1}(p_\perp,p_1) 
\partial_H B(p_\perp,p_1)\right),
\end{equation}
where the lower case symbol tr denotes a partial trace with respect to
the discrete indices $k$, $l$, and the matrix $B$ is given by
$B_{kl}(p_\perp,p_1)=N_{k-l}(p_\perp,p_1+2\pi l/\lambda)$ with the
$N_m$ defined by Eq.~(\ref{eq:decomp}). In the following, we will
apply this formula to the profile of Fig.~\ref{fig:1}(b) for which the
matrices $N_m$ can be obtained analytically.\cite{Emig03} 

For $\lambda \to 0$, i.e., if $\lambda$ is the smallest length scale,
the function $g(p_\perp,p_1)$ can be calculated analytically, leading
to $g(p_\perp,p_1)=2p/(e^{2p(H-a)}-1)$ for both TM and TE modes. After
insertion into Eq.~(\ref{eq:F-non-pert}), one finds the simple result
\begin{equation}
\label{eq:f-rd}
F_{\rm 0} = - \frac{\pi^2}{480} \frac{\hbar c}{(H-a)^4}.
\end{equation}
for the force from TM or TE modes, respectively. This result can be
obviously interpreted as the force between two planar plates at a
reduced distance $H-a$ which reflects that the relevant modes do not
"feel" the narrow valleys of the profile in the limit $\lambda \ll
a,H-a$. In the opposite limit of very large $\lambda$, the PA should
be reliable, with the result 
\begin{equation}
\label{eq:f-rect-PA}
F_{\rm PA} = -\frac{\pi^2}{480}\frac{\hbar c}{2} 
\left(\frac{1}{(H+a)^4}+\frac{1}{(H-a)^4}\right),
\end{equation}
for both TM and TE modes. Here the PA yields the same result for
the flat and the corrugated surface based approximation.

\begin{figure}[h]
\begin{center}
\epsfxsize=28pc 
\epsfbox{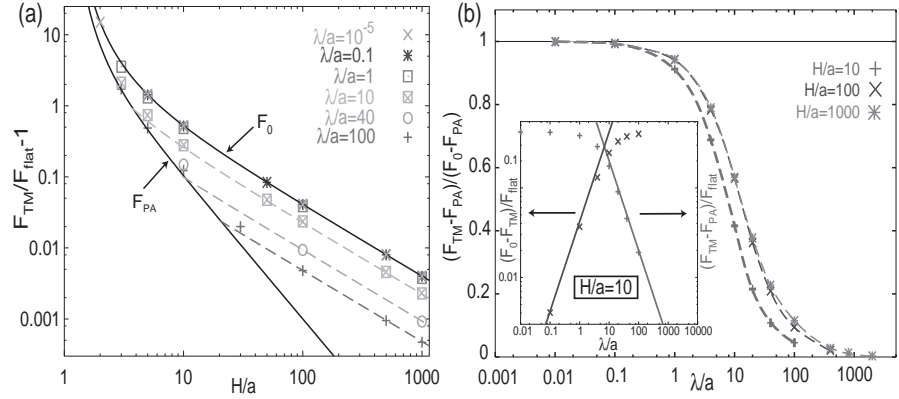} 
\caption{Casimir force $F_{\rm TM}$ from TM modes for the geometry of Fig.~\ref{fig:1}(b). (a) Relative change compared to the force between two flat plates for different corrugation lengths $\lambda$. (b) Crossover between the proximity approximation at large $\lambda$ [Eq.~(\ref{eq:f-rect-PA})] and the reduced distance result of Eq.~(\ref{eq:f-rd}) at small $\lambda$.  \label{fig:2}}
\end{center}
\end{figure}

For all intermediate values of $\lambda$, the force can be calculated
numerically from Eq.~(\ref{eq:F-non-pert}) by defining the $M$-th
order approximation $g_M(p_\perp,p_1)$ to the function
$g(p_\perp,p_1)$ by truncating the matrix $B_{kl}$ symmetrically
around $(k,l)=(0,0)$ at order $M$ so that the trace in
Eq.~(\ref{eq:fct-g}) extends only over $k$,
$l=-(M-1)/2,\ldots,(M-1)/2$. This in turn, by numerical integration of
Eq.~(\ref{eq:F-non-pert}), yields a series of forces which can be
extrapolated to the actual force at $M\to\infty$. The results for TM
modes have been obtained in Ref.~\cite{Emig03} and are shown in
Fig.~\ref{fig:2}. The force $F_{\rm TM}$ at fixed $H/a$ is found to be
monotonous in $\lambda/a$ and its minimal value is given by the PA
result $F_{\rm PA}$ of Eq.~(\ref{eq:f-rect-PA}). At small $\lambda/a$
the exact result of Eq.~(\ref{eq:f-rd}) is recovered, and the
divergence seen in perturbation theory, Eq.~(\ref{eq:F-small-lam}), is
removed.  However, in agreement with perturbation theory for the
sinusoidal profile, the relative change $F_{\rm TM}/F_{\rm flat}-1$ of
the force shows a crossover at $\lambda \simeq H$ from a $H^{-2}$
decay at $H\lesssim \lambda$ to a $H^{-1}$ decay at larger $H$.
Fig.~\ref{fig:2}(b) shows how the limits of Eqs.~(\ref{eq:f-rd}),
(\ref{eq:f-rect-PA}) are approached with decreasing or increasing
$\lambda$ at fixed $H/a$. In the range of studied values for $H/a$ the
crossover between the two limits appears at $\lambda/a$ of order $10$.
As shown by the inset of Fig.~\ref{fig:2}(b) for $H/a=10$, we find
that the corrections to the results of Eqs.~(\ref{eq:f-rd}),
(\ref{eq:f-rect-PA}) obey power laws,
\begin{equation}
  \label{eq:scaling}
  \frac{F_{\rm 0}-F_{\rm TM}}{F_{\rm flat}} \sim \frac{\lambda}{a}, \quad
  \frac{F_{\rm TM}-F_{\rm PA}}{F_{\rm flat}} \sim \frac{a}{\lambda}
\end{equation}
for $\lambda/a \ll 1$ and $\lambda/a \gg 1$, respectively. For TE
modes we expect in the first case a decay $\sim (\lambda/a)^{1/2}$
while in the latter case the exponent will be the same as for TM
modes. Comparing to Eq.~(\ref{eq:diff-to-PA}) we observe that the PA
limit is approached with a different exponent for large $\lambda$.
This scaling behavior can be understood from classical ray optics.
Following a recently proposed "optimal" PA for a scalar
field\cite{Jaffe+03}, the Casimir energy is not obtained from the
normal distance based on one of the surfaces [cf.~Eq.~(\ref{eq:pa})]
but from the {\it shortest} surface-to-surface ray (of length
$\ell(\by_\|,z)$) through given positions $(\by_\|,z)$ between the
two surfaces of area $A$,
\begin{equation}
  \label{eq:PA-opti}
  {\cal E}_{\rm opt}/{\cal E}_0= \int d^2 \by_\| \int_{h(\by_\|)}^H dz\,
\frac{H^3}{A \, \ell^4(\by_\|,z)}.
\end{equation}
Of course, this is still an approximation even for small surface
curvature since the surfaces are treated as locally flat.  However, we
expect that this approximation yields the correct scaling behavior for
large $\lambda/a$. For the rectangular corrugation the lengths of the
shortest paths can be computed analytically. For small $a/H$ the
dominant deviation from the standard PA comes from rays through
positions located in equal sectors of almost triangular cross section
for which the rays end exactly on the upper edges of the corrugated
profile. For large $\lambda$ the volume of a sector becomes {\it
  independent} of $\lambda$.  In this case, a simple calculation
yields, using Eq.~(\ref{eq:PA-opti}), the correction to the PA force,
$(F_{\rm opt}-F_{\rm PA})/F_{\rm flat} = \frac{112}{9} \sqrt{a/H}
a/\lambda$, which reproduces our numerically observed decay $\sim
a/\lambda$ of Eq.~(\ref{eq:scaling}).  How can this be reconciled with
our perturbative result of Eq.~(\ref{eq:diff-to-PA}) for the
sinusoidal profile? The crucial point is that the latter profile has a
finite slope across the entire surface. Thus, the shortest length
$\ell$ deviates for {\it all} positions between the surfaces from the
distance used in the standard PA. Using Eq.~(\ref{eq:PA-opti}), a
model calculation for a corrugation with a piecewise constant slope
indeed reproduces a decay $\sim (a/\lambda)^2$ of the correction to
the standard PA result.

\begin{figure}[h]
\begin{center}
\epsfxsize=23pc 
\epsfbox{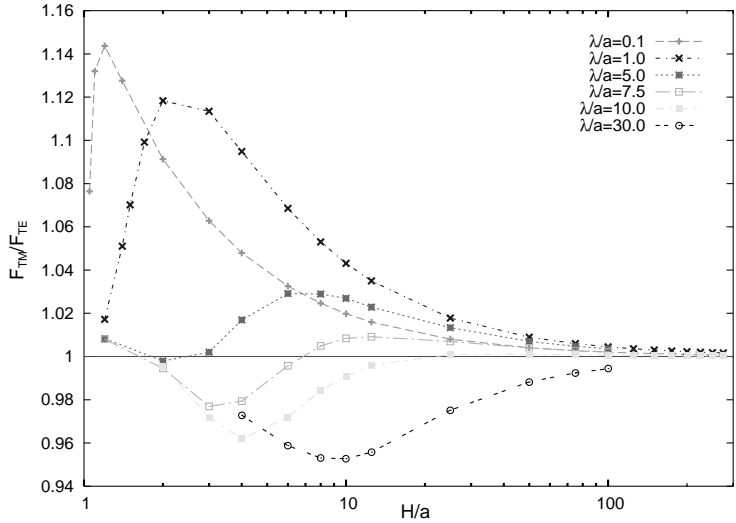} 
\caption{Ratio of the Casimir force from TM and TE modes, respectively, for the geometry of Fig.~\ref{fig:1}(b). \label{fig:3}}
\end{center}
\end{figure}

So far, we presented numerical results for TM modes only. For TE modes
the same approach can be applied to the matrix of
Eq.~(\ref{eq:kernels}). The full electrodynamic Casimir force is then
obtained as $F_{\rm TM}+F_{\rm TE}$.  Instead of presenting the
results for TE modes independently, it is more useful to look at their
relative contribution compared to TM modes. Fig.~\ref{fig:3} displays
the ratio $F_{\rm TM}/F_{\rm TE}$ of the force from TM and TE modes
for different corrugation lengths. Since the force is again bounded
from below by $F_{\rm PA}$ and from above by $F_0$, one has $F_{\rm
  TM}/F_{\rm TE} \to 1$ for both $\lambda/a\to\infty$ and
$\lambda/a\to 0$. However, for intermediate corrugation lengths the
ratio can differ from one, with the amount depending on the mean
surface distance $H$. We observe the tendency that TM mode
contributions dominate at smaller $\lambda$ while for larger $\lambda$
the TE modes provide a larger force. In contrast, for a sinusoidal
profile the force from TM modes is larger at all $\lambda$, at least
to second order in $a$.  For very small separations $H \gtrsim a$ the
segments of the rectangular corrugation which are closer to the flat
surface yield the main contribution and the PA is expected to hold,
implying $F_{\rm TM}/F_{\rm TE}\approx 1$ which indeed can be observed
in Fig.~\ref{fig:3}. In the opposite case of large $H \gg a$, one
obtains the flat plate geometry, and again $F_{\rm TM}/F_{\rm TE}\to
1$.  The latter limit is approached for both types of corrugations
with a power law $\sim (H/a)^{-1}$ which we find in perturbation theory
as well as in the numerical approach.

\section*{Acknowledgments}
I thank B. Duplantier for useful discussions and R. Golestanian, A.
Hanke and M. Kardar for their collaboration on part of the work
presented here. This project is supported by the Deutsche
Forschungsgemeinschaft (DFG) through an Emmy Noether grant.

\end{document}